\documentclass[twocolumn,prl,preprintnumbers,amsmath,amssymb]{revtex4}

\usepackage{graphicx}
\usepackage{dcolumn}
\usepackage{bm}
\usepackage{mathrsfs}
\usepackage{color}
\usepackage{epstopdf}
\usepackage{booktabs}

\newcommand{\ba}{\begin{array}}
\newcommand{\ea}{\end{array}}
\newcommand{\be}{\begin{equation}}
\newcommand{\ee}{\end{equation}}
\newcommand{\bea}{\begin{eqnarray}}
\newcommand{\eea}{\end{eqnarray}}

\begin{document}

\title{\fontsize{11,5}{\baselineskip}\selectfont Tunable high-resolution macroscopic self-engineered geometric phase optical elements}


\author{Etienne Brasselet*}

\affiliation{Univ. Bordeaux, CNRS, LOMA, UMR 5798, F-33400 Talence, France}

\date{\today}

\begin{abstract}
Artificially engineered geometric phase optical elements may have tunable photonic functionalities owing to sensitivity to external fields, as is the case for liquid crystals based devices. However, a liquid crystal technology combining high-resolution topological ordering with tunable spectral behavior remains elusive. Here, by using a magneto-electric external stimulus, we create robust and efficient self-engineered liquid crystal geometric phase vortex masks with broadly tunable operating wavelength, centimeter-size clear aperture, and high-quality topological ordering.
\end{abstract}

\maketitle

Two decades ago Bhandari proposed the use of space-variant anisotropic structures to design flat optical elements imparting a polarization-controlled wavefront to a paraxial incident field \cite{bhandari_physrep_1997}. The idea consists to consider a slab of inhomogeneous uniaxial material that locally behave at a conventional uniform half-wave retarder. As it passes through the optical element, an incident circularly polarized light field thus reverses its handedness. The geometry of the system also leaves its imprint as the polarization and spatial degrees of freedom couples one to another during propagation---a process that is referred to as the spin-orbit interaction for light \cite{bliokh_natphot_2015}. More precisely, the output light acquires an extra phase term $\exp[2i\sigma\psi(r,\phi)]$ where $\sigma=\pm1$ is the photon helicity that refers to the circular polarization state handedness and $\psi(r,\phi)$ is the in-plane optical axis orientation angle, $r$ and $\phi$ being the polar coordinates in the plane of the slab. Such helicity-dependent phase term has a purely geometric nature, it is the Pancharatnam-Berry geometric phase associated with rotated coordinate systems. Geometric phase physics is a longstanding concept \cite{berry_phystoday_1990} that received a substantial attention in optics, especially when dealing with singular optics that is, quoting Soskin and Vasnetsov who introduced the terminology, ``this new branch of modern physical optics deals with a wide class of effects associated with phase singularities in wave fields, as well as with the topology of wave fronts''  \cite{soskin_progopt_2001}. Indeed, geometric phase optical elements are nowadays commonly used to produce light beams carrying on-axis optical phase singularities---vortex beams---associated to an amplitude factor $\exp(i\ell\phi)$, where the integer topological charge $\ell$ is controlled by the photon helicity. In practice, this is made by using azimuthally varying half-wave retarders characterized by $\psi = q \phi$ where $q$ is a half-integer or an integer, which gives $\ell = 2\sigma q$.

Geometric phase optical vortex generators were initially fabricated using subwavelength gratings in the mid-infrared \cite{biener_ol_2002}. They became widely used once implemented in the visible domain using liquid crystals \cite{marrucci_prl_2006} and later extended to other media as well, such as liquid crystal polymers \cite{mceldowney_ol_2008} or glasses \cite{beresna_apl_2011}. To date, handy macroscopic `q-plates' all rely on machining techniques to either pattern the optical axis of naturally birefringent materials or create artificial birefringence on-demand. In addition, there are independent strategies to tune the operating wavelength of these elements via birefringence control under external fields \cite{piccirillo_apl_2010, slussarenko_oe_2011} and to fabricate high-resolution static topological ordering of the optical anisotropy \cite{tabirian_ieee_2015}, irrespective of the value of $q$. However, macroscopic standalone optical elements combining all these assets are still missing, which is the motivation of the present work.

In this work we propose to reach the unprecedented technical specifications mentioned above by taking advantage of the self-organization of liquid crystals molecules around topological defects in thin planar films (i.e., thickness of the order of ten micrometers). The main idea is to direct the spontaneous topological structuring of liquid crystals by applying an external orientational stimuli in presence of degenerate orientational boundary conditions at the input and output facets of the film. The spontaneous generation liquid crystal topological defects has been already explored in various environments and external fields \cite{voloschenko_ol_2000, brasselet_prl_2009, brasselet_ol_2011, barboza_prl_2012, loussert_prl_2013, pieranski_lc_13, budagovsky_Bull_2015}, however, none of them meet the sought-after specifications. Here, by using initially homogeneous nematic liquid crystal films under the influence of a magneto-electric external field, we produce self-engineered liquid crystal q-plates with electrically tunable operating wavelength, centimeter-size clear aperture and high-quality topological ordering. In turn, spectrally agile and pure optical vortex generation using collimated macroscopic laser beam is demonstrated while the high-resolution features are illustrated in the framework of spiral phase Fourier filtering.

\begin{figure}[t!]
\centering
\includegraphics[width=1\columnwidth]{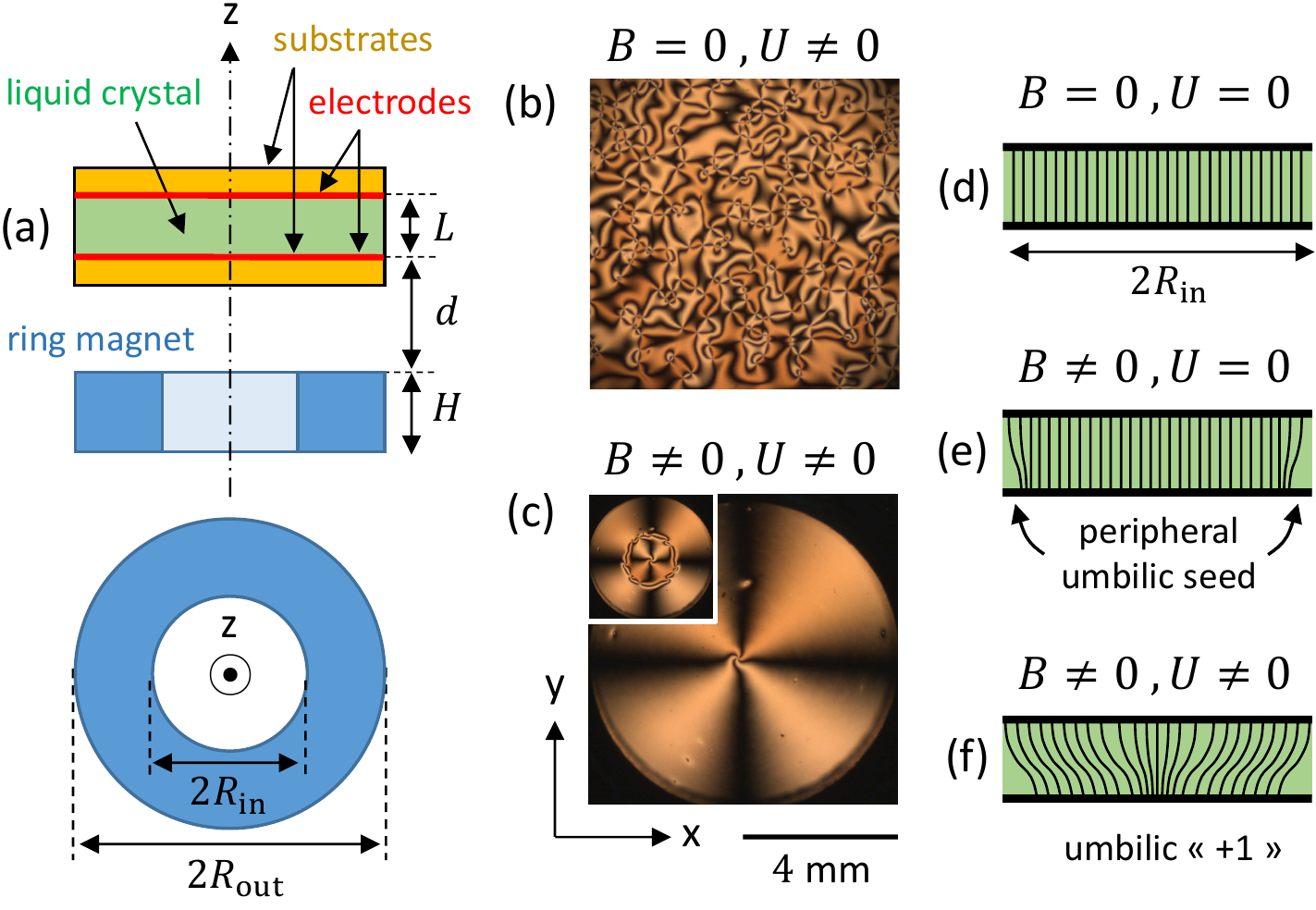}
\caption{
(a) Geometry of the liquid crystal film under magneto-electric external field. The magnetization of the Nickel-plated Neodynium (grade N42) ring magnet is directed along the $z$ axis. The magnet parameters are $R_{\rm in}=4~$mm, $R_{\rm out}=7.5~$mm and $H=6~$mm, and is associated with a pull force of $\sim 50~N$ (manufacturer datasheet). (b,c) Images of the sample observed between crossed linear polarizers (oriented along the $x$ and $y$ axes) using white light incoherent illumination 10~min after 	a square-waveform voltage $U=2.15$~V at 2~kHz frequency is applied without (b) and with (c) the presence of the ring magnet. The magnet is placed at a distance $d=2$~mm from the liquid crystal layer in panel (c) while $d=1$~mm in  inset. The dark cross reveals an umbilical defect with topological charge $+1$. (d,e,f) Side view of the lines of the director field---the local-averaged orientation of the liquid crystal molecules---of the sample at rest (d), under the magnetic field alone (e) and under the combined action of the magnetic and electric field (f).
}
\end{figure}

Our experimental arrangement is shown in Fig.~1(a). The sample is a film of nematic liquid crystal with thickness $L=20~\mu$m, birefringence $\delta n = 0.097$ at 589~nm wavelength and negative dielectric anisotropy $\epsilon_a = -6.4$ at 1~kHz frequency, both evaluated at $25^\circ$C, and positive magnetic anisotropy (yet not measured). The liquid crystal slab is sandwiched between two glass substrates provided with transparent electrodes and perpendicular orientation for the liquid crystal molecules. At rest, the liquid crystal orientation is therefore homogeneous and aligned along the $z$ axis, as depicted in Fig.~1(d). When a uniform quasi-static electric field is applied alone, a collection of umbilical defects with topological charges $\pm1$ appears randomly above a threshold applied voltage and undergoes long-term collective annihilation dynamics, as early unveiled by Rapini \cite{rapini_jphys_1973}. The latter situation is shown in Fig.~1(b), where the observation of the sample between crossed linear polarizers reveals a random network of defects that are individually identified as extinction crosses. On the other hand, a single defect with topological charge $+1$ can be obtained under a uniform magnetic field alone, as shown by Pieransky and coworkers \cite{pieranski_lc_13} when using a disk-magnet. However, this approach provides neither with clear aperture nor almost uniform birefringent phase retardation. These problems are solved by using the combined action of the static magnetic field from an annular permanent magnet placed at a distance $d$ from the liquid crystal layer, see Fig.~1(a). In that case, the magnetic field alone seeds the clear aperture of the annular magnet with a peripheral umbilical liquid crystal structure, as depicted in Fig.~1(e), which arises from the thresholdless magnetic torque associated with the axisymmetric transverse component of the magnetic field. Although the corresponding liquid crystal reorientation is hardly perceived between crossed linear polarizers, the magnetic topological structuring is revealed when combined with the effect of a uniform electric field. Indeed, this eventually leads to a macroscopic steady umbilic with topological charge $+1$ centered on-axis as shown experimentally in Fig.~1(c) for $d=2$~mm and depicted in 1(f). Note that the distance $d$ between magnet at the liquid crystal plays a role in the nature of the steady topological structure, as illustrated in the inset of Fig.~1(c) for which $d=1$~mm, which we attribute to the axial reversal of magnetic field lines \cite{lebedev_tm_1991} that are known to lead to inversion walls for the liquid crystal orientation as previously shown in another geometry \cite{gilli_jphys_1994}. For magnetic field spatial distribution, one refers to Ref.~\citenum{ravaud_ieee_2008}. In contrast the axisymmetry of the resulting structure is progressively lost at large $d$, in agreement with the random reorientation without magnetic field shown in Fig.~1(b). In what follows we thus fix $d=2$~mm, which offers uniform retardance over the full clear aperture of the device, except nearby the defect core where it rapidly vanishes over a characteristic distance (the core radius) that decreases as the  voltage increases, as is the case for random umbilics created by an electric field alone \cite{rapini_jphys_1973}.

\begin{figure}[b!]
\centering
\includegraphics[width=\columnwidth]{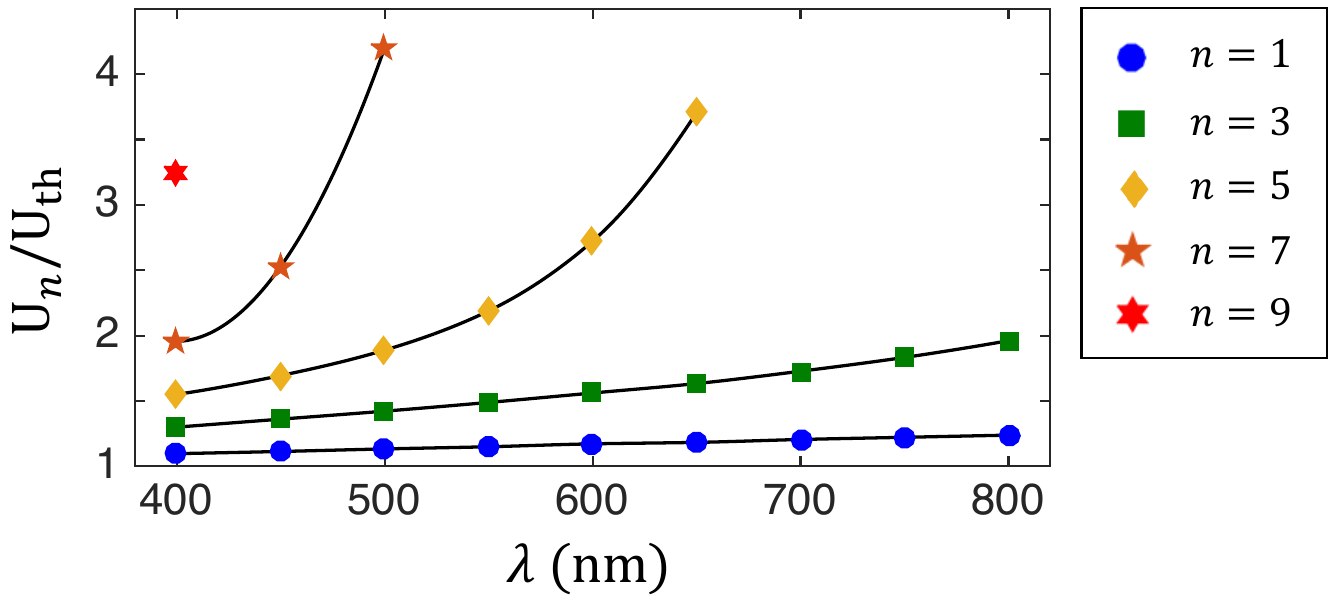}
\caption{
Experimental spectral dependence of the reduced voltage values $U_n/U_{\rm th}$, where $U_{\rm th} \simeq 1.8~$V refers to the threshold below which the liquid crystal remains unperturbed, that fulfill the half-wave retardance condition $\Delta_\infty = n\pi$ with $n$ odd integer. Color markers refer to different values for $n$ while the black curves are only guide for the eyes.
}
\end{figure}

\begin{figure*}[t!]
\centering
\includegraphics[width=2\columnwidth]{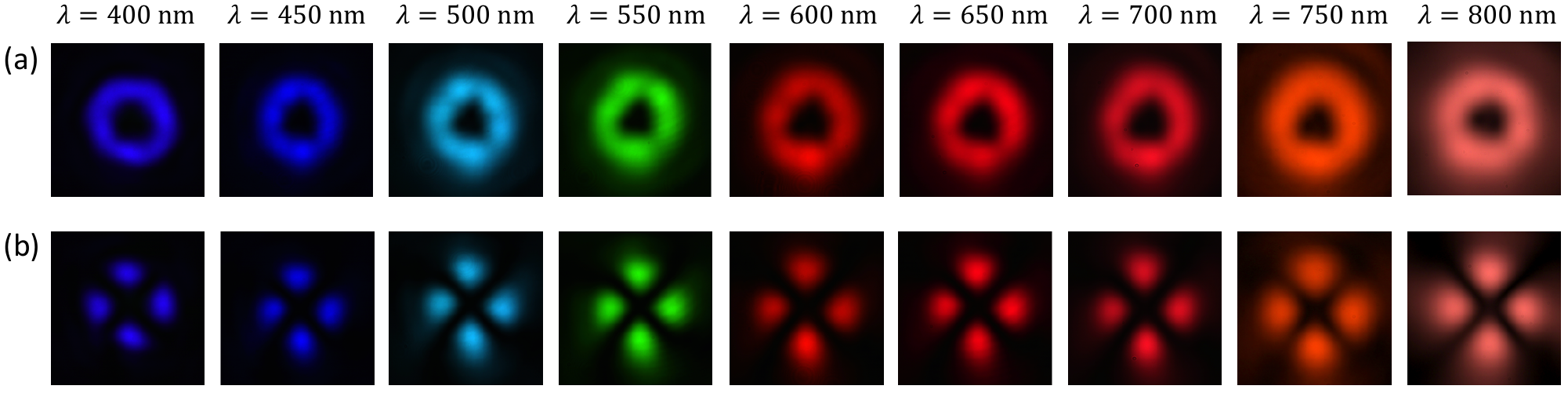}
\caption{
Optical characterization of the generation of spectrally agile optical vortex beams from a 2~mm diameter collimated supercontinuum laser beam filtered with a set of nine wavelengths from 400~nm to 800~nm by steps of 50~nm, in the case of an incident linear polarization state along the $x$ axis. The study is performed for $n=1$, hence adjusting the applied voltage according to the specifications reported in Fig.~2. (a) Far-field total intensity profiles. (b) Far-field intensity patterns when a linear polarizer oriented along the $x$ axis is placed at the output of the device.
}
\end{figure*}

Our magneto-electric umbilic thus behaves as an electrically tunable q-plate with $q=+1$ where the asymptotic (i.e., outside the core region) birefringent phase retardation $\Delta_\infty$ can be typically tuned between 0 and $2\pi \delta n L/\lambda$. Therefore the half-wave retardation condition can be satisfied up to the wavelength $\lambda_{\rm max} = 2\delta n L$, which corresponds to the mid-infrared domain in the present case. For shorter wavelength, the half-wave condition can be fulfilled for $m>1$ values of the voltage for as long as the asymptotic retardance can reach $\Delta_\infty = (2m-1)\pi$. This is experimentally illustrated in Fig.~2 that displays the voltage values $U_n$ associated with $\Delta_\infty = n\pi$ ($n$ being an odd integer) in the full visible domain by using a collimated supercontinuum laser beam with 2~mm diameter impinging at normal incidence on the sample along the $z$ axis. The measurements are made by assessing the voltage values maximizing the transmitted power by placing the sample between crossed circular polarizers. Data is collected for a set of nine wavelengths in steps of 50~nm, with 10~nm full-width at half-maximum transmission spectrum. Accordingly, the demonstration of spectrally agile optical vortex generation at the macroscopic scale is illustrated in Fig.~3 for a zero-order q-plate (i.e., $n=1$) in the particular case of an incident linear polarization state along the $x$ axis. Recalling that a linear polarization corresponds to an equal-weight coherent superposition of orthogonal circular polarization states, the output beam consists of the superposition of contra-circularly polarized scalar vortex beams with opposite topological charges $\pm2$. Namely, a so-called vectorial vortex beam often referred to as a nonseparable optical state. As expected, annular-shaped total intensity profiles beams are obtained as shown in Fig.~3(a), while the vectorial underlying structure is revealed by placing a linear polarizer at the output of the sample, see Fig.~3(b). Indeed, maximally contrasted four-lobe patterns are observed in the latter case according to the expected inhomogeneous linear polarization state whose azimuth rotates by $4\pi$ over a full turn.

The proposed magneto-electric strategy thus enables the fabrication of a self-engineered vectorial vortex mask with large clear aperture that performs well over a broad spectral range without need of post-polarization filtering as is usually required for moderately pure geometric phase optical vortex generators. Moreover, this method is robust regarding the long-term stability of the defect whose location marginally evolves as shown in Fig.~4(a), which corresponds to rms radial displacement of the order of $1.5~\mu$m over up to $\sim 20$ hours with neither temperature nor mechanical vibration control. The magneto-umbilic is also resilient to moderately-high optical power, as demonstrated in Figs.~4(b), 4(c) and 4(d) that respectively illustrate barely noticeable structural changes as the optical power of a 2~mm-diameter incident circularly polarized Gaussian beam at 532~nm wavelength increases up to 1~W. From a practical point of view, we note that the temperature range of the device is dictated by the existence of the nematic phase, which basically depends on the used material. Since the nematic phase exists in the present case from $-40^\circ$C to $+101^\circ$C we deal here with a very robust situation where the temperature can be safely varied by up to several tens of degrees from room temperature conditions. Of course, since dielectric and elastic properties are temperature-dependent the half-wave retardation condition is met at slightly different voltages values as the temperature varies while the q-plate behavior is preserved. 

\begin{figure}[b!]
\centering
\includegraphics[width=\columnwidth]{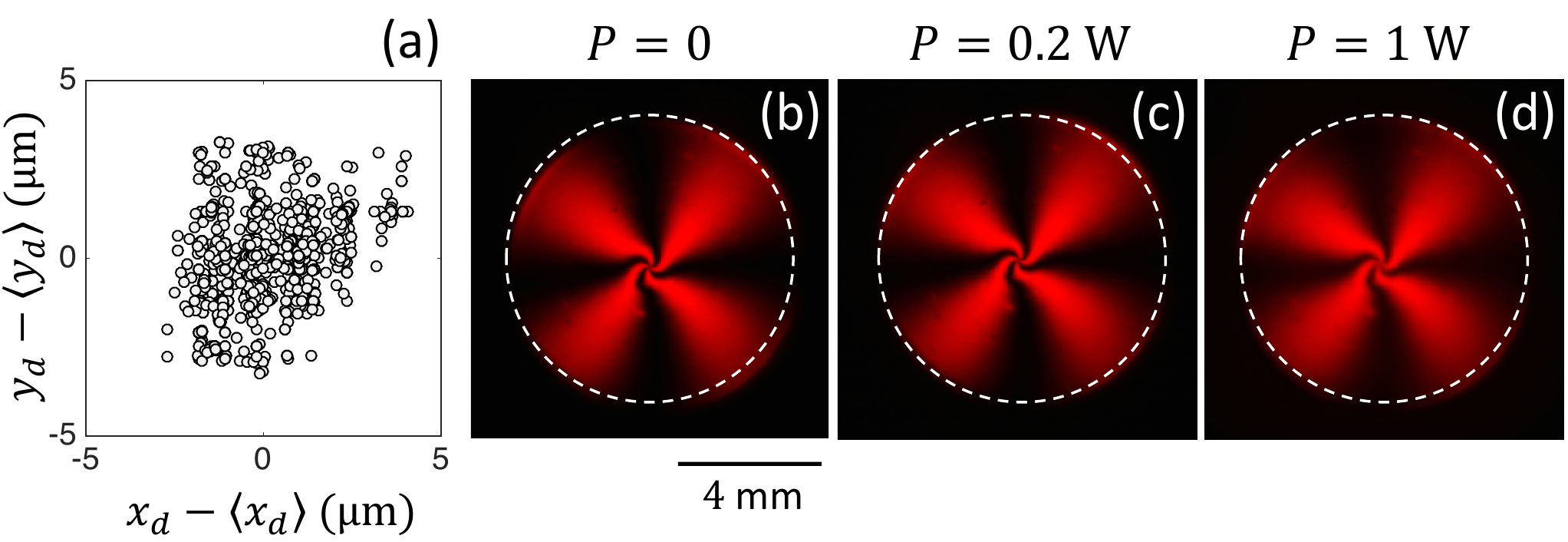}
\caption{
(a) Temporal evolution of the defect location given by its coordinates $(x_d, y_d)$ in the plane of the film over almost 20 hours, where $\langle \cdot \rangle$ denotes time average over the whole duration of the observation. (b,c,d) Imaging of the device placed between crossed linear polarizers using white light incoherent illumination spectrally filtered at 633~nm wavelength when a circularly polarized Gaussian beam at 532~nm wavelength with total optical power $P$ impinges at normal incidence onto the sample. Shown data recorded at $U/U_{\rm th}=1.2$.}
\end{figure}

In addition, two independent structural features deserve comment. First, the in-plane distribution of the director field exhibits an axisymmetric twisted director field, as already shown in Figs.~1 and 4. This implies an effective optical axis orientation angle of the form $\psi(r,\phi) = \phi+\varphi(r)$ where the nonzero swirl function $\varphi(r)$ is a generic consequence of the elastic anisotropy of the liquid crystal, as already discussed in the particular case of liquid crystal light valves \cite{barboza_ptrsa_2014}. Noteworthy, the swirl only imparts smooth changes to the output wavefront curvature via an pure-phase amplitude term $\exp[2i\sigma\varphi(r)]$, which therefore let the optical vortex generation process unaltered. Still, it is interesting to recall that the latter helicity-dependent curvature can be either a drawback \cite{kravets_pra_2018} or an asset \cite{tam_prappl_2017} when dealing with nonseparable optical states. The fact that the applied voltage is a control parameter for the swirl, as illustrated in Fig.~5, thus appears as an interesting additional degree of freedom. Second, the core area can be electrically controlled as is the case for random umbilics \cite{rapini_jphys_1973}, as illustrated in Fig.~5(b), which gives a recipe to optimize the uniformity of the optical element by reducing the area with nonuniform retardance, especially when optical processing is desirable at the microscopic scale as discussed hereafter.

\begin{figure}[t!]
\centering
\includegraphics[width=1\columnwidth]{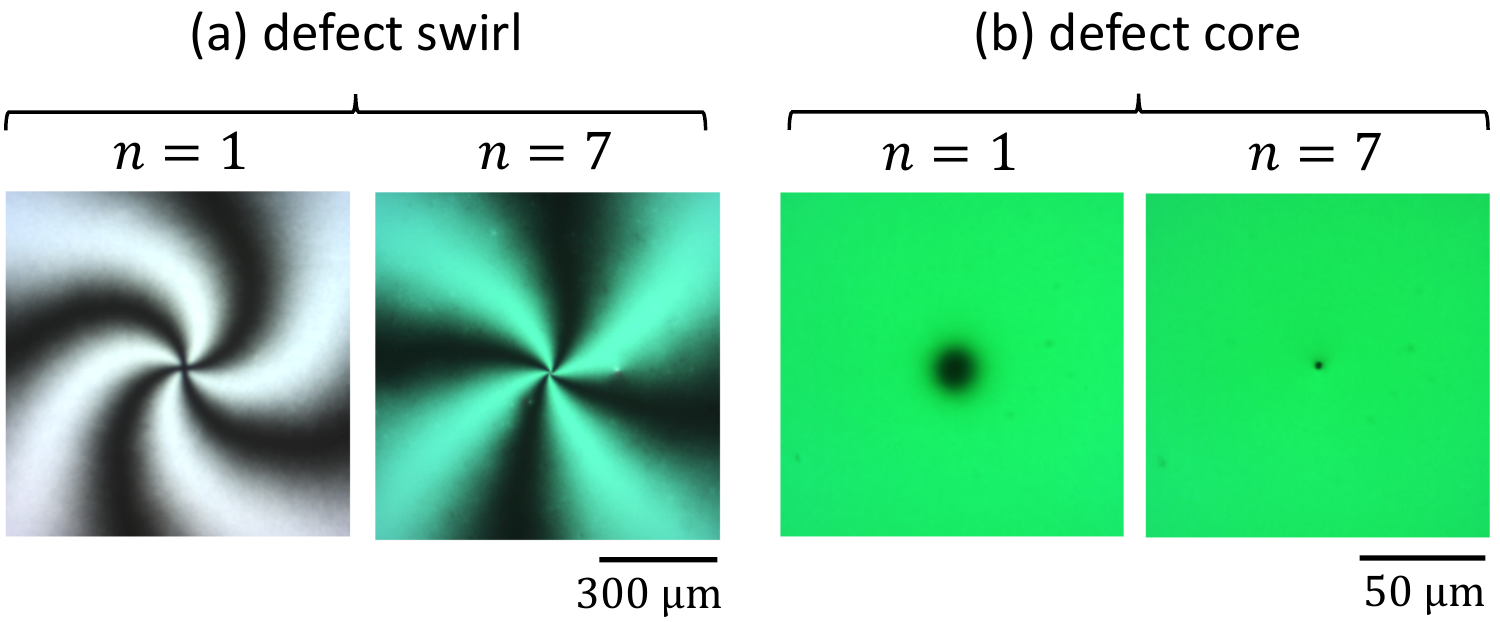}
\caption{
(a) Observation of the swirl in the neighborhood of the liquid crystal topological defect by placing the device between crossed linear polarizers using white light incoherent illumination. (b) Observation of the umbilic core by placing the sample between crossed circular polarizers under white light incoherent illumination spectrally filtered at 532~nm wavelength. Shown data refer to applied voltage values that correspond to $n=1$ and $n=7$ for 532~nm wavelength.}
\end{figure}

This happens for optical imaging techniques relying on spiral phase Fourier filtering towards resolution and sensitivity enhancement. This has been implemented in various standard and super-resolution microscopy techniques or for astronomical imaging, as reviewed in \cite{ritsch_ptrsa_2017}. Here we illustrate it in a case corresponding to the basic scheme used in optical vortex coronagrahy where phase masks producing optical vortices with even topological charges are used \cite{swartzlander_jo_2009}. In practice, this is done by placing the sample in the Fourier plane of a telescope and reimaging an input clear circular aperture according to the set-up sketched in Fig.~6(a). If the defect is not centered on the zeroth-order Fourier spot, the image of the aperture is almost unperturbed, as shown in Fig.~6(b). In contrast, when the defect is placed on-axis, the light intensity is redistributed at the periphery of the geometric image of the input pupil for odd $n$, as shown in Fig.~6(c). Ideally, the diffracted light can be totally suppressed by placing a diaphragm in the exit input pupil plane, which lead to the coronagraphic effect in the context of astronomical imaging of faint celestial objects nearby bright sources of light. We emphasize that no post polarization filtering is used, in contrast to previous attempts using self-engineered small-scale liquid crystal q-plates \cite{aleksanyan_ol_2016, aleksanyan_prl_2017}. In principle, one should also benefit from thinner liquid crystal layers since the core radius scales as $L$ \cite{rapini_jphys_1973}, though a tradeoff is expected recalling the effect of the `topological magnetic seed' is driven by the reorienting magnetic torque that scales as $L^{-2}$. Nevertheless, the quest for ever smaller umbilic core radius is not a necessary condition for potential applications. Indeed, previous works showed that scalar vortex phase masks without phase singularity in the central area are useful to obtain shadow effects in spiral phase contrast imaging \cite{jesacher_prl_2005} and vectorial vortex phase masks without uniform birefringent phase retardation in the central area can generate optical vortex arrays from a single defect\cite{brasselet_prl_2012}.

\begin{figure}[t!]
\centering
\includegraphics[width=0.95\columnwidth]{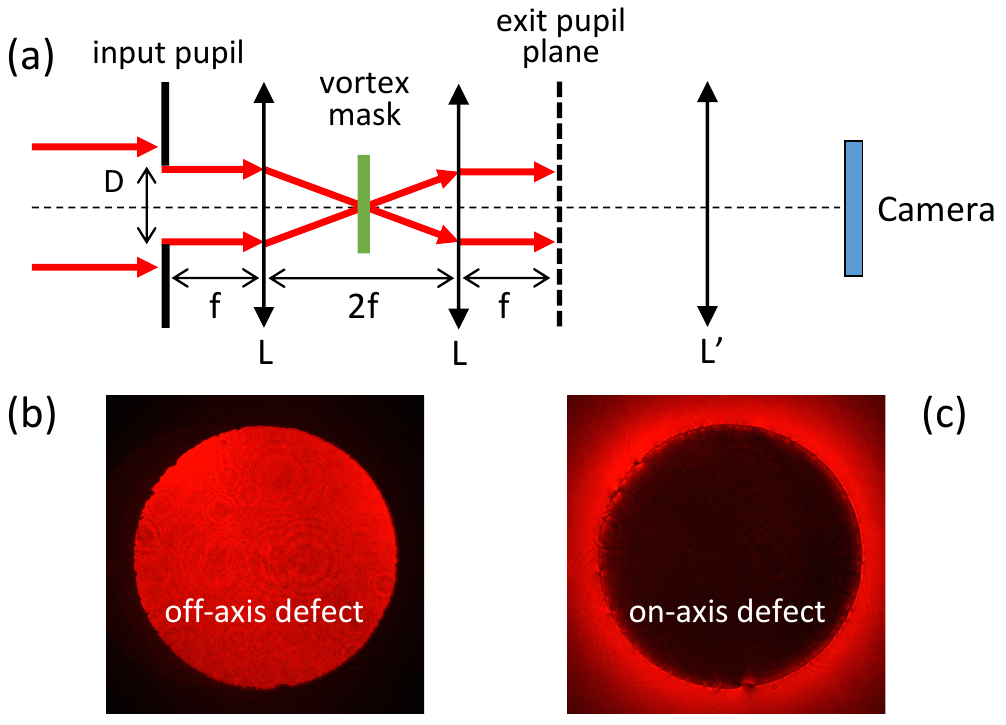}
\caption{
Experimental set-up used for the demonstration of spiral phase Fourier filtering at 700~nm wavelength using a so called $4f$ optical system made of a pair of lenses L with focal length 200~mm. This corresponds to an Airy spot radius $r_0 \simeq 40~\mu$m (taken at the first zero of intensity) in the Fourier plane of the telescope. The input circular aperture has a diameter $D=4$~mm. The self-engineered magneto-electric liquid crystal q-plate is placed in the Fourier plane. The input pupil is reimaged on a camera using a lens with focal lens 75~mm when the defect is displaced off-axis at a distance $\simeq 10 r_0$ (b) and when it is placed on-axis (c), for $n=5$.
}
\end{figure}

Summarizing, the combined action of a static magnetic with a quasi-static electric field enables the robust self-engineering of handy macroscopic liquid crystal q-plates. Large clear aperture of the order of 1~cm$^2$, high-quality topological structuring at small scale ensured by a nonsingular topological defect for the director field, and an operating wavelength that is electrically tunable over an ultra-broadband spectral range are obtained. The proposed technology-free and Nature-assisted approach to create high-tech geometric phase optical elements invite to further consider soft-matter self-organization processes in spin-orbit photonics technologies.
\vspace{2mm}

This work has been partially funded by the CNRS program ``D{\'e}fi Instrumentation aux limites 2017''.
\vspace{2mm}

*etienne.brasselet@u-bordeaux.fr


\end{document}